\documentclass[10pt,letterpaper,twocolumn]{article} 

\usepackage{ol2}
\usepackage[draft]{hyperref}
\usepackage{amsmath}

\begin{document}

\twocolumn[ 

\title{Low phase noise  diode laser oscillator for 1S\,--\,2S spectroscopy in atomic hydrogen}


\author{N.~Kolachevsky,$^{1,2}$ J.~Alnis,$^{1}$ C.G.~Parthey,$^{1}$ A.~Matveev,$^{1}$ R.~Landig,$^{3}$,  T.W.~H\"{a}nsch$^{1,4}$}

\address{
$^1$Max-Planck-Institut f\"{u}r Quantenoptik, 85748 Garching,
Germany
\\
$^2$P.N. Lebedev Physics Institute, 119991 Moscow,
Russia \\
$^3$Institute for Quantum Electronics, ETH Z\"urich, CH-8039 Z\"urich, Switzerland\\
$^4$Ludwig-Maximilians-Universit\"at M\"unchen, 80799 M\"unchen,
Germany}

\begin{abstract}We report on a low-noise diode laser oscillator at 972\,nm actively
stabilized to an ultra-stable vibrationally- and thermally
compensated reference cavity. To increase the fraction of laser
power in the carrier we designed a 20\,cm long external cavity
diode laser with an intra-cavity electro-optical modulator. The
fractional power in the carrier reaches 99.9\,\% which corresponds
to a rms phase noise of
$\varphi^2_\textrm{rms}=1\,\textrm{mrad}^2$ in 10\,MHz bandwidth.
Using this oscillator we recorded 1S\,--\,2S spectra in atomic
hydrogen and have not observed any significant loss of the
excitation efficiency due to phase noise multiplication in the
three consecutive 2-photon processes.
\end{abstract}


 ] 

\noindent Frequency stabilized diode lasers have become a powerful
tool for high-resolution spectroscopy and precise optical
frequency measurements \cite{Margolis1995, Baillard2007}.
The combination of wide tuning range, compact size, low power
consumption and reasonable price make them excellent light sources
for e.g. transportable frequency standards \cite{Kersten2004}.
Significant progress in stability and size reduction is reached by
implementation of  mid-plane suspended reference cavities
\cite{Notcutt2005}. On the other hand, semiconductor power
amplifiers allow efficient generation of higher harmonics which
can be used for excitation of narrow atomic transitions in the UV
region \cite{Kolachevsky2006}. In this case, requirements to the
frequency stability of an oscillator become much more stringent.
First, each two-photon process quadruples the spectral line width
of an oscillator emitting a phase-diffusion field \cite{Ryan1995}.
Secondly, the multiplication of phase noises in a multi-photon
process results in a reduction of the power accumulated in the
carrier and losses in the excitation efficiency of narrow atomic
resonances \cite{Telle1996}. The latter is a serious problem for
diode lasers because of their excessive noise level
\cite{Henry1982}. The spectral line width of a solitary laser
diode may reach hundreds of megahertz while optical and/or
electronic feedback can suppress phase fluctuations only to a
certain extent.

Previously \cite{Kolachevsky2006,Alnis2008} we reported on a
frequency-quadrupled master-oscillator power-amplifier (MOPA)
diode laser system  for high-resolution two-photon 1S\,--\,2S
spectroscopy in atomic hydrogen at 243 nm. The presence of
short-correlated phase noise of the oscillator reduced the
excitation efficiency to about 40\% compared to the case if the
whole power was concentrated in the carrier. This paper describes
design and characterization of a low phase noise  diode oscillator
at 972\,nm  which has already been used in a number of precision
measurements providing the significant increase of their accuracy
\cite{Kolachevsky2009,Parthey2010}. Besides applications in
high-resolution spectroscopy of hydrogen and, probably, of
anti-hydrogen \cite{Andresen10, Amoretti02, Gabrielse02}, this
laser system may be used for creating an intense cold beam of
metastable hydrogen atoms. The latter may improve the accuracy of
spectroscopy of highly excited states in H \cite{Biraben09} and
facilitate resolving the proton charge radius puzzle
\cite{Pohl2010}.

\begin{figure}[b!]
\centerline{
\includegraphics[width=8cm]{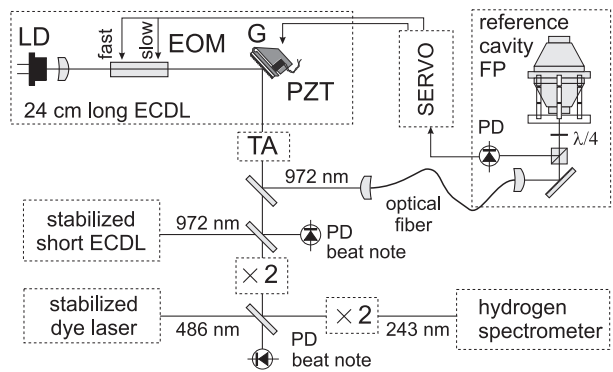}}
 \caption{\small Schematic of the setup. EOM --
electro-optical modulator; G -- diffraction grating; PZT -- piezo
actuator; TA -- tapered amplifier; PD -- photodiode. The long ECDL
is actively stabilized to the vibrationally and thermally
compensated ULE cavity FP \cite{Alnis2008}.}\label{fig1}
\end{figure}

The key improvement was the design of the external cavity diode
laser (ECDL) in a Littrow configuration at 972\,nm with a cavity
length of $L=20$\,cm long (see Fig.\,\ref{fig1}). According to
\cite{Riehle2004}, reduction of the phase noise and narrowing of
the (unlocked) ECDL spectral line width is achieved by increasing
the length of the external cavity according to
$\Delta\nu=\Delta\nu_\textrm{LD}/(1+L/(nL_\textrm{LD}))^2$, where
$\Delta\nu_\textrm{LD}$ is the spectral line width of the solitary
diode, $n$ and $L_\textrm{LD}$ are its refractive index and the
length. Other approaches  \cite{Doringshoff2007} may be used as
well.

Our long ECDL consists of  a 100\,mW  laser diode (without
anti-reflection coating), an electro-optical modulator (EOM, model
PM25 from Linos) and a 1200\,lines/mm holographic diffraction
grating with a diffraction efficiency of 30\%. The single mode
operation was readily achieved by adjustment of the injection
current while the typical mode hop-free tuning range is of
$500$\,MHz. The ECDL is mounted on a temperature-controlled
 breadboard placed in a sealed metal box additionally
protected from outside by a heavy plywood box. This setup allows
to find the desired wavelength in a few minutes and provides
robust operation in an air-conditioned laboratory over the whole
day.

Similar to \cite{Stoehr2006,Muller2006} we used a Brewster angle
intra-cavity EOM to control the ECDL frequency.  An attempt to use
the injection current modulation resulted in increased phase noise
and caused mode hops. To suppress noise coming from injection
current we filter it by a chain of capacitors and inductances
directly at the laser diode. Fine rotation and tilt adjustments of
the EOM are necessary to minimize the amplitude modulation which
strongly influences the lock quality.

\begin{figure}[t!]
\centerline{
\includegraphics[width=6cm]{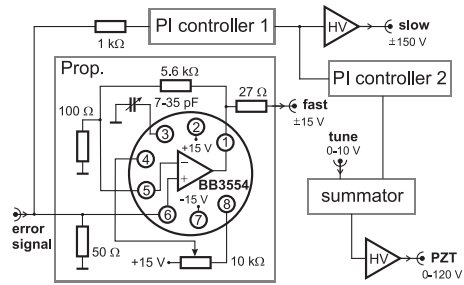}}
 \caption{\small Schematics of the servo loop used to control the frequency of the long cavity ECDL.
 Notations of output signals correspond to  Fig.\,\ref{fig1}. The error signal is taken directly
 from the output of the phase detector RPD-1 (Mini-Circuits) heterodyning the PDH photodiode output. PI denote conventional proportional-integrating
controllers, HV --- high voltage amplifiers. }\label{fig2}
\end{figure}

The laser is phase-stabilized with the help of the
Pound-Drever-Hall (PDH) method \cite{Drever1983} to the
vibrationally and thermally compensated Ultra Low Expansion (ULE)
glass Fabry-P\'erot cavity  described in details in
\cite{Alnis2008}. Electronics used for locking the laser is
sketched in Fig.\,\ref{fig2}. For compensation of fast
fluctuations a fast proportional BB3554-based amplifier with
3\,MHz bandwidth is connected to one of the EOM's electrodes.
Robust long-term locking is  achieved using a
proportional-integrating controller (PI\,1) and the high-voltage
amplifier connected to the second electrode of the EOM. Even
better long-term lock stability is reached by the second
controller (PI\,2) working as pure integrator with a time constant
of 1\,s used to stabilize the average output of PI\,1 via
controlling the voltage on the piezo-element attached to the
grating. Such combination provides stable and robust lock over
more than 5 hours.

The ULE cavity has a symmetrical configuration  which
significantly suppresses the influence of vertical accelerations
\cite{Notcutt2005} while low sensitivity to thermal fluctuations
is provided by cooling the cavity to the ULE zero expansion point
of $+12.5\,^\circ$C using Peltier elements \cite{Alnis2008}. This
design provides sub-hertz spectral line width of the laser at a
drift rate of about 50\,mHz/s.

To characterize the phase noise of the ECDL we recorded the beat
note of its second harmonic  with a stabilized 486\,nm dye laser
\cite{Fischer2004}. Although the carrier of the dye laser is much
broader than that of the diode laser (60\,Hz compared to 1\,Hz at
486\,nm), it is a useful tool to study short-correlated phase
fluctuations of the diode laser. The dominating phase
perturbations in the dye laser come from acoustic fluctuations in
the dye jet and have much lower frequencies compared to ones in
the diode lasers.

\begin{figure}[t!]
\begin{center}
\includegraphics [width=6cm]{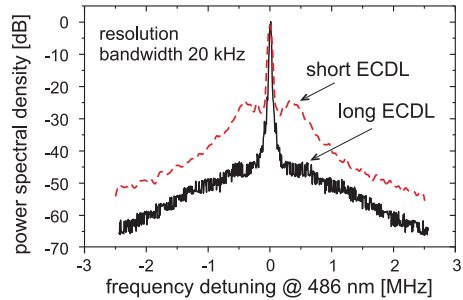}
\caption{(Color online) {\small Power spectrum density of the beat
notes between the stabilized dye laser and the second harmonic of
the long  ECDL (solid line) or the short ECDL (dashed line).
Carrier amplitudes are normalized to 0\,dB.}}\label{fig3}
\end{center}
\end{figure}

The beat note spectrum  is shown in Fig.\,\ref{fig3}. We ascribe
the  noise pedestal  around the carrier to the ECDL since it shows
some sensitivity to the parameters of ECDL's electronic feedback
and is independent of the dye laser electronics. When we switch
off the feedback of the long ECDL, the spectrum shape remains
almost the same which indicates that the free running ECDL has a
line width of less than 20\,kHz. Calculating the power fraction in
the noise pedestal, we evaluate  the rms phase noise of the long
ECDL
 of $\varphi^2_\textrm{rms}=1\,\textrm{mrad}^2$ at 972 nm in
10\,MHz bandwidth.

Using the same method we investigated the phase noise of a short
972\,nm ECDL with a cavity length of 2\,cm
\cite{Kolachevsky2006,Alnis2008}. The laser is locked to the
second vibrationally and thermally compensated ULE cavity with the
same characteristics as the first one. For this laser we could
achieve the lowest rms phase noise of only
$\phi^2_\textrm{rms}=13\,\textrm{mrad}^2$ (10\,MHz bandwidth)
 by a fine adjustment of the electronic feedback. It
was a compromise between the light power coupled to the cavity,
bandwidth and gain of the servo loop.

\begin{figure}[t!]
\begin{center}
\includegraphics [width=6cm]{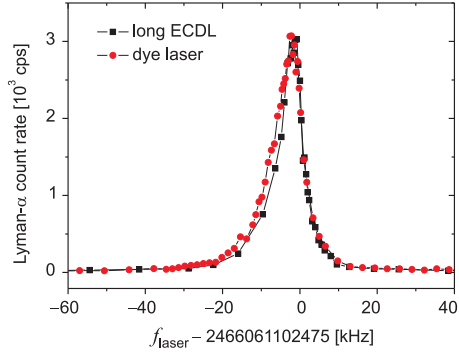}
\caption{(Color online) {\small 1S\,--\,2S spectrum of atomic
hydrogen recorded with the 20\,cm long ECDL (squares) and dye
laser (circles). The power levels of light at 243\,nm exciting the
two-photon transition were equal for both cases. The broader
spectrum recorded using the dye laser is the result of its broader
line width.}
 }\label{fig4}
\end{center}
\end{figure}

When amplifying and frequency quadrupling the long ECDL, up to
15\,mW of spectrally pure light at 243\,nm become available for
hydrogen spectroscopy. The direct comparison between 1S\,--\,2S
excitation rates for the long ECDL oscillator based system and the
frequency doubled dye laser is shown in Fig.\,\ref{fig4}. Atoms
excited from the ground state in the cold beam at 13\,K are
detected by counting Lyman-$\alpha$ photons which are emitted when
the 2S state decays in an electric field (for details see e.g.
\cite{Parthey2010}). Recording the spectra in a few minutes
interval we have not observed any significant difference in the
excitation efficiencies.

 Since 8 photons at 972 nm contribute to an excitation of the 2S state,
for the long ECDL we expect the change  of an excitation
efficiency by a factor of only  $\eta=\exp[-(8
\varphi_\textrm{rms})^2]=0.94$ \cite{Telle1996} which turned out
to be rather insensitive  on adjustments of the electronic
feedback. For comparison using the short ECDL one could reach an
excitation efficiency of only $\eta=0.44$, which, together with a
high sensitivity to feedback parameters, did not make it suitable
for high-precision experiments.

\begin{figure}[t!]
\begin{center}
\includegraphics [width=6cm]{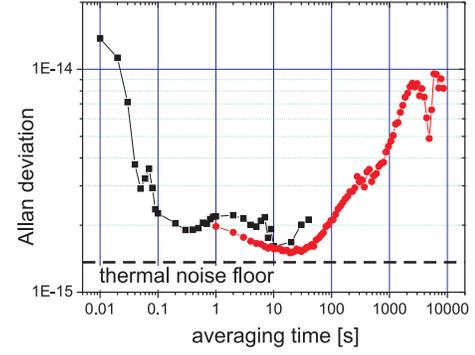}
\caption{\small{(Color online) The Allan deviation of the beat
note frequency between the long and the short ECDL's stabilized to
two equivalent independent ULE cavities. Drift is not subtracted.
Data are recorded by FX80 Klische+Kramer counter with the gate
times of 10\,ms (squares) and 1\,s (circles).}}\label{fig5}
\end{center}
\end{figure}

Along with excellent short-correlated phase noise characteristics,
long ECDL possesses high long-term stability of the temperature
and vibrationally compensated ULE  cavity. Fig.\,\ref{fig5} shows
an Allan deviation plot recorded for the beat note between the
long and the short  ECDLs  stabilized to two independent but
similar ULE cavities. The Allan deviation nearly reaches the
thermal noise floor evaluated as $1.4\times10^{-15}$ for the
cavities \cite{Numata2004}.

In conclusion, we have developed a diode laser system at 972 nm
with a low phase noise level of
$\varphi^2_\textrm{rms}=1\,\textrm{mrad}^2$ in 10\,MHz bandwidth
and 0.5\,Hz spectrally narrow carrier containing 99.9\% of the
laser power. This compact system allows efficient excitation of
the
 $1S$\,--\,$2S$ transition in H and D, which is useful for high precision experiments
 and production of metastable H beams.

The work is supported by Munich Advanced Photonics and Marie Curie
actions. NK acknowledges support from Presidential Grant No.
MD-669.2011.8 (Russian Federation).


\begin{thebibliography}{99}


\bibitem{Margolis1995} H.S.\,Margolis, G.P.Barwood, G.\,Huang, H.A.\,Klein, S.N.\,Lea, K.\,Szymaniec, and P.\,Gill, {\it Science} {\bf 306}, 1355 (2004).
%
\bibitem{Baillard2007} M.M.\,Boyd, A.D.\,Ludlow, S.\,Blatt, S.M.\,Foreman, T.\,Ido, T.\,Zelevinsky, and
J.\,Ye, {\it Phys. Rev. Lett.} {\bf 98}, 083002 (2007).
%
\bibitem{Kersten2004} P.\,Kersten, F.\,Mensing, U.\,Sterr,
F.\,Riehle, {\it Appl. Phys.} B {\bf 68}, 27 (2004).
%
\bibitem{Notcutt2005} M.\,Notcutt, L.-S.\,Ma, J.\,Ye, and J.L.\,Hall., {\it Opt. Lett.} {\bf30}, 1815
(2005).
%
\bibitem{Kolachevsky2006} N.\,Kolachevsky, J.\,Alnis, S.D.\,Bergeson, and T.W.\,H\"ansch, {\it Phys. Rev.} A {\bf 73}, 021801(R)
(2006).
%
\bibitem{Ryan1995} R.E.\,Ryan, L.A.\,Westling, R.\,Bl\"umel,
H.J.\,Metcalf, {\it Phys. Rev.} A {\bf 52}, 3157 (1995).
%
\bibitem{Telle1996} H.R.\,Telle, in: {\it Frequency Control of Semiconductor Lasers}, ed. by M. Ohtsu, 137, Wiley, New York,
(1996).
%
\bibitem{Henry1982} C.H.\,Henry, {\it IEEE J.Quantum Electron.} {\bf
QE-18}, 259 (1982).
%
\bibitem{Alnis2008} J.\,Alnis, A.\,Matveev, N.\,Kolachevsky,
T.\,Udem, and T.W.\,H\"ansch, {\it Phys. Rev.} A, {\bf 77}, 053809
(2008).
%
\bibitem{Kolachevsky2009} N.\,Kolachevsky {\it et al.}, {\it Phys.
Rev. Lett.} {\bf 102}, 213002 (2009).
%
\bibitem{Parthey2010} C.G.\,Parthey {\it et al.}, {\it Phys.
Rev. Lett.} {\bf 104}, 233001 (2010).
%
\bibitem{Andresen10} Andresen {\em et al.}, {\it Nature} {\bf 468}, 673 (2010).
\bibitem{Amoretti02} Amoretti {\em et al.}, {\it Nature} {\bf 419}, 456 (2002).
\bibitem{Gabrielse02} Gabrielse {\em et al.}, {\it Phys. Rev. Lett.} {\bf 89}, 213401 (2002).
%
\bibitem{Biraben09} F.\,Biraben, {\it Eur. Phys. J. Special Topics}  {\bf 172}, 109
(2009).
%
\bibitem{Pohl2010} R.\,Pohl {\it et al.}, {\it Nature} (London),
{\bf 466}, 213 (2010).
%
\bibitem{Riehle2004} F.\,Riehle, Frequency standards, 295, Wiley VCH Verlag, Weinheim (2004).
%
%
\bibitem{Doringshoff2007} K.\,D\"oringshoff, I.\,Ernsting,
R.-H.\,Rinkleff, S.\,Shiller, and A.\,Wicht, Opt. Lett. {\bf 32},
2876 (2007).

%
\bibitem{Stoehr2006} H.\,Stoehr, F.\,Mensing, J.\,Helmke, and
U.\,Sterr, Opt. Lett. {\bf 31}, 736 (2006).
%
%
\bibitem{Muller2006} H.\,M\"{u}ller, S.-W.\,Chiow, Q.\,Long, and S.\,Chu, Opt. Lett. {\bf 31}, 202
(2006).
%
\bibitem{Drever1983} R.W.P.\,Drever, J.L.\,Hall, F.V.\,Kowalski,
J.\,Hough, G.M.\,Ford, A.J.\,Munley, and H.\,Ward, Appl. Phys. B
{\bf 31}, 97 (1983).
%
%
\bibitem{Fischer2004} M. Fischer {\it et al.}, Phys. Rev. Lett. {\bf 92}, 230802
(2004).


\bibitem{Numata2004} K.\,Numata, A.\,Kemery, J.\,Camp, Phys. Rev. Lett. {\bf 93}, 250602
(2004).




\end{thebibliography}
\end{document}